\documentclass{elsart}
\usepackage{multicol}
\usepackage[dvips]{graphics}

\begin{document}

\begin{frontmatter}
  \title{Structure and diffusion time scales of disordered clusters}

  \author{E. Cuansing\corauthref{cor}}, \ead{cuansing@physics.purdue.edu} 
  \author{H. Nakanishi}
  \corauth[cor]{Corresponding author.}
  \address{Department of Physics, Purdue University, West Lafayette,
           IN 47907-1396}

  \begin{abstract}
    The eigenvalue spectra of the transition probability matrix
    for random walks traversing critically disordered clusters
    in three different types of percolation problems show that
    the random walker sees a developing Euclidean signature
    for short time scales as the local, full-coordination constraint
    is iteratively applied.
  \end{abstract}


\end{frontmatter}

Physical processes such as diffusion in porous media, electrical and
thermal conduction in composite materials, and information flow in 
random networks may be modeled using random walks on disordered 
clusters.  For random walks on uniform Euclidean systems, their 
mean-square displacement goes as $<R^2(t)>\ \sim\ t$, where $t$ is the 
number of steps.  In contrast, constraining the random walker to 
traverse a critically disordered cluster such as an incipient 
infinite percolation cluster makes its mean-square displacement go as 
$<R^2(t)>\ \sim\ t^{2/d_w}$, where $d_w > 2$ is known as the walk 
dimension.  This slowing down of the random walker, called 
{\em anomalous diffusion} (see, for reviews, 
\cite{havlin87,nakanishi94}), is due to the delays it experiences 
by being trapped in or reflected from the irregular features of the
cluster.

The disordered cluster where the random walker is constrained to traverse 
can be of many different types.  For this article, we have chosen 
to investigate three closely-related models.  These models are ordinary 
\cite{stauffer94}, fully coordinated (FC) \cite{cuansing99}, and iterated 
fully coordinated (IFC) percolation \cite{cuansing01}, all on the square 
lattice.  Although the disordered clusters produced by each of these
models look qualitatively different, it has been found previously 
that they exhibit the same static and dynamic critical 
properties \cite{cuansing01}.  This work extends the analysis by 
investigating the entire eigenvalue spectra of their corresponding 
transition probability matrices \cite{nakanishi94}.  Furthermore, we have 
investigated the effects of culling either fully-coordinated (FC) sites 
or partially-coordinated (PC) sites on the structure of the
disordered clusters and the characteristics of the random walk on them.

\begin{figure}
  \centering{\resizebox{4.5in}{4.5in}{\includegraphics{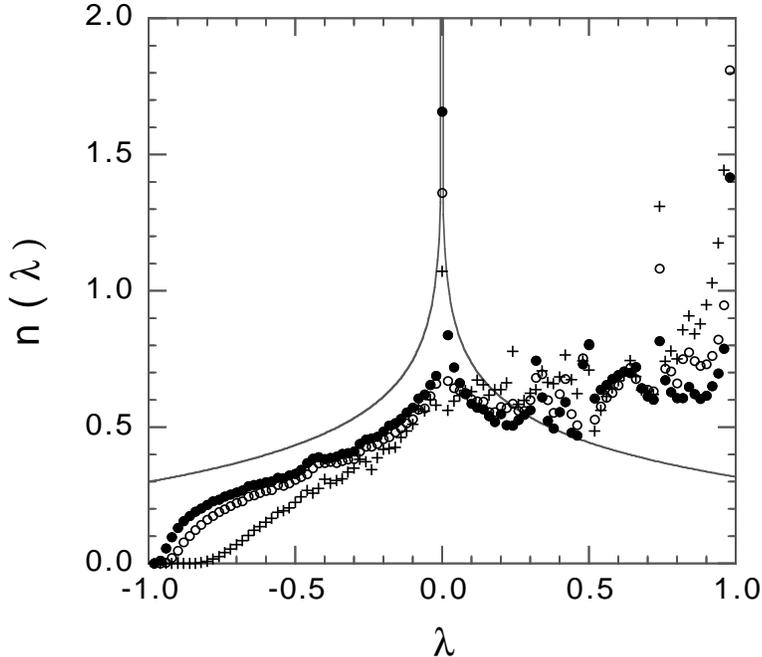}}}
  \vspace*{-1.2in}
  \caption{Plot of the eigenvalue spectra for ordinary $(+)$,
           FC $(\circ)$, and IFC $(\bullet)$ percolation.
           Each set of data points is obtained from $1200$ lattice 
           realizations of size $128 \times 128$.  The solid curve 
           shown is the smoothed plot of the eigenvalue spectrum for a 
           uniform Euclidean system of size $120 \times 120$.}
  \vspace*{0.2in}
  \label{fig:spec}
\end{figure}

The eigenvalues $\lambda$ of the transition probability matrix $W$ directly
provide the time-dependent behavior of the random walker 
\cite{nakanishi93,mukherjee94}.  Eigenvalues near $\lambda = 1$ characterize
the long time behavior of the random walker while those at the region near
$\lambda = 0$ only enter the characterization of the random walker's short
time behavior.

In Fig.~\ref{fig:spec} the eigenvalue spectra for a blind ant random 
walker traversing a critically disordered cluster constructed from 
ordinary, FC, and IFC percolation are shown together with the sketch for 
the eigenvalue spectrum for a random walk on a uniform Euclidean lattice.
At the region near $\lambda = 1$, all three percolation models exhibit the 
same power-law behavior.  Quantitatively, the eigenvalue density 
$n(\lambda)$ near $\lambda = 1$ scales according to 
$n(\lambda)\ \sim\ |\ln \lambda|^{d_s/2 - 1}$ \cite{nakanishi93}.  For 
the above three models, therefore, the long time behavior of the random 
walker is the same.  This agrees with the previous result that the three 
models exhibit the same walk dimension $d_w$ \cite{cuansing01}.

\begin{figure}
\begin{multicols}{2}
  \centering{\resizebox{2.58in}{2.28in}{\includegraphics{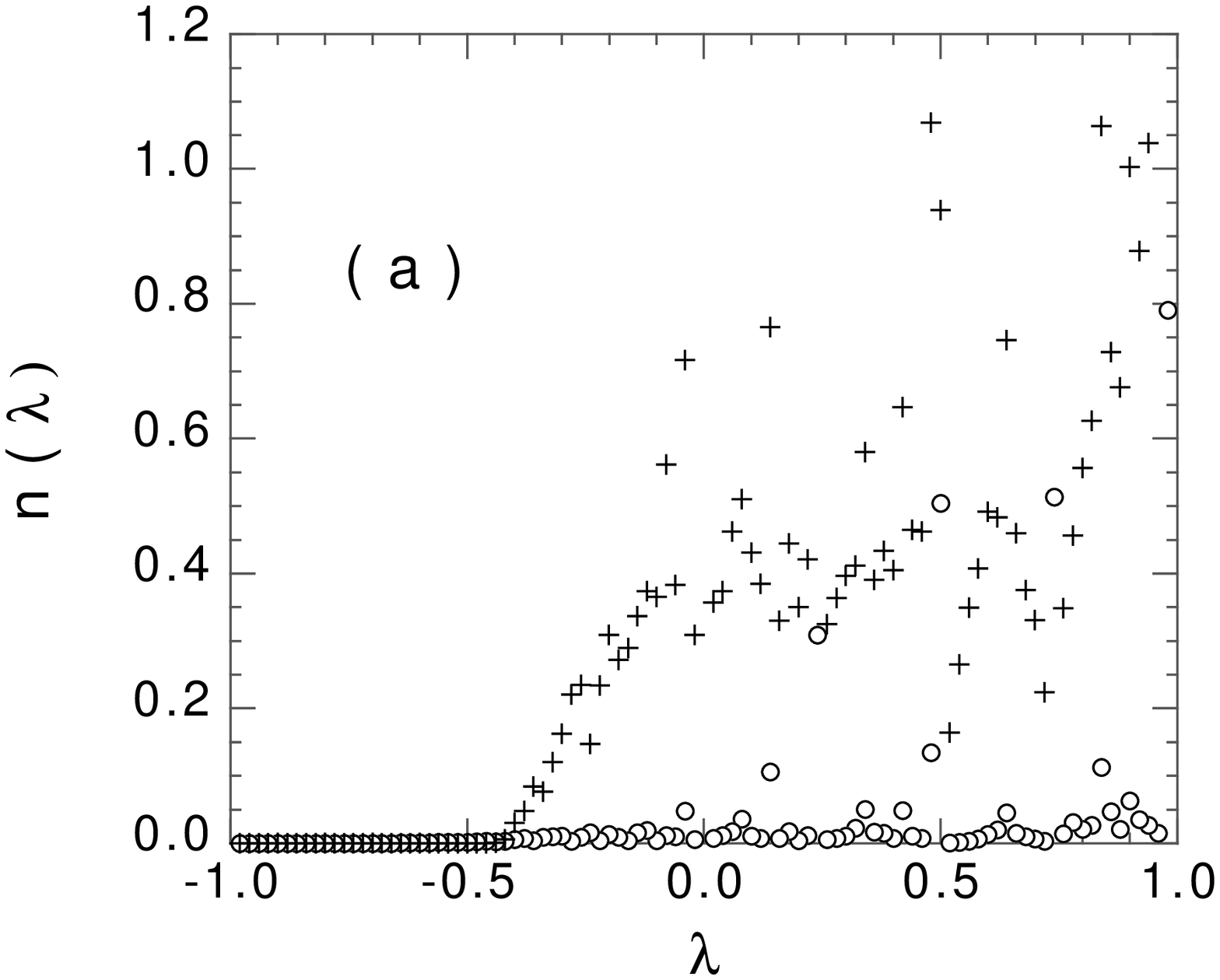}}}\\
  \centering{\resizebox{2.58in}{2.28in}{\includegraphics{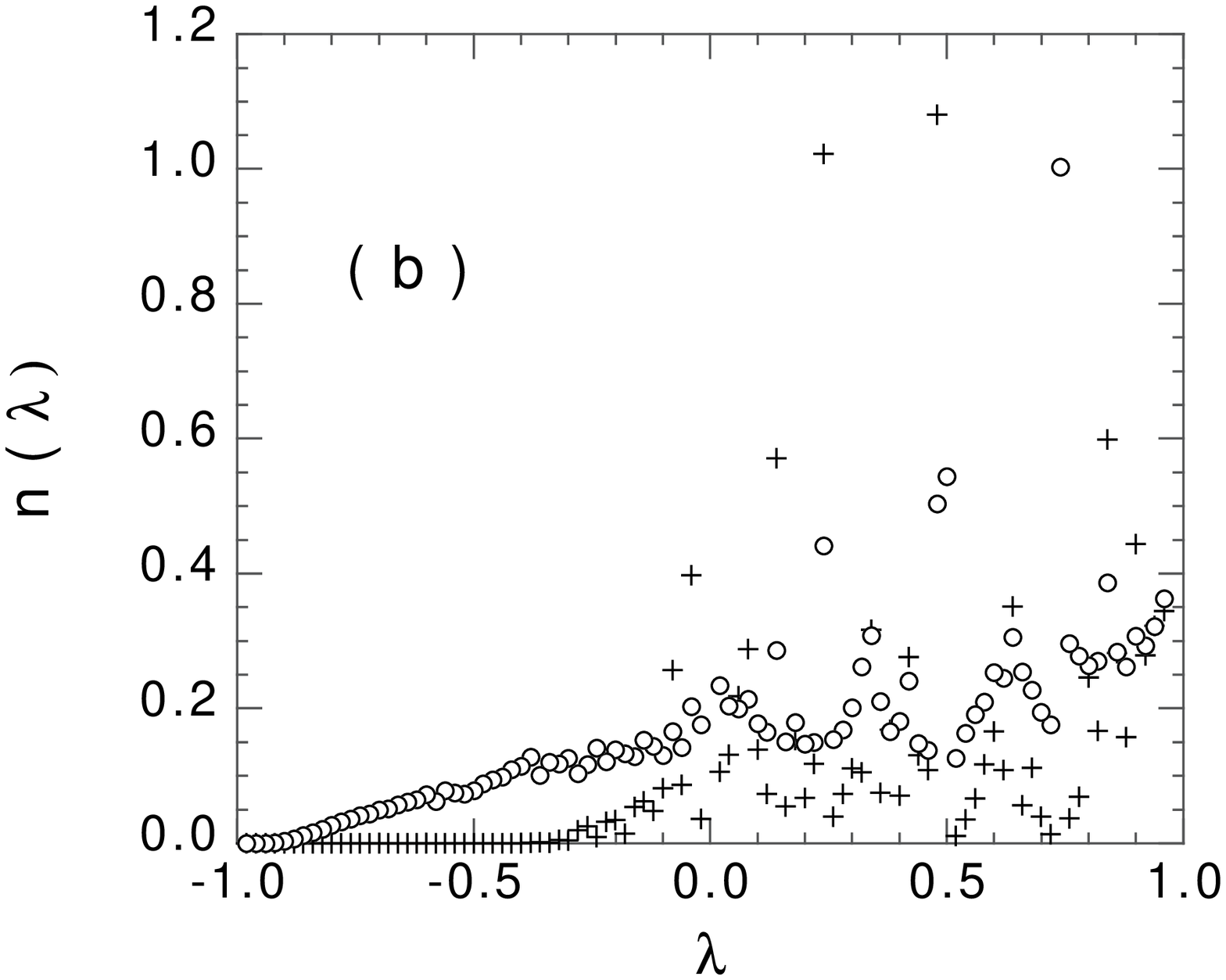}}}
\end{multicols}
\centering{\resizebox{2.58in}{2.28in}{\includegraphics{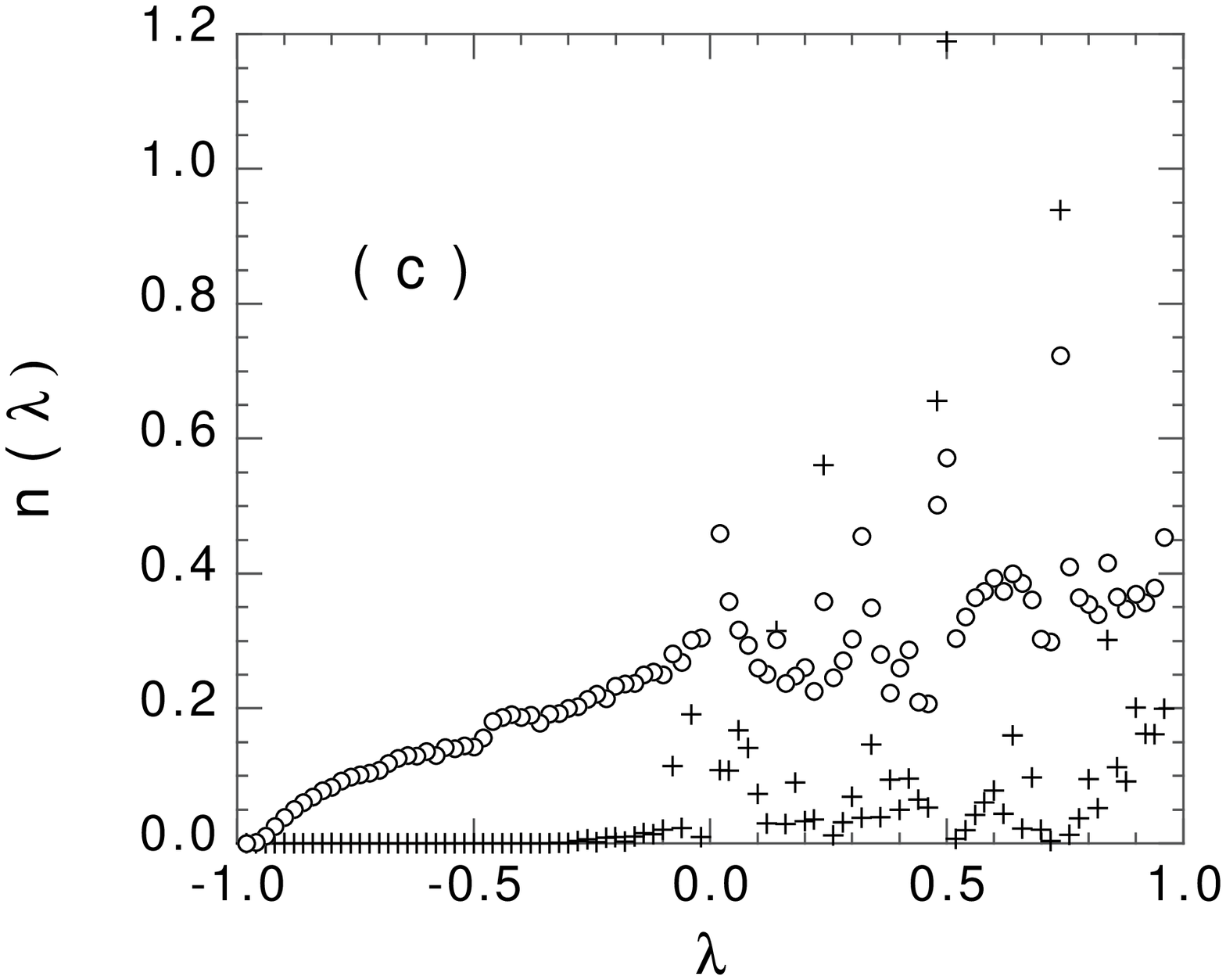}}}
\caption{Eigenvalue spectra for (a) ordinary, (b) FC, and (c) IFC percolation,
         when FC and PC sites are culled.  For (a), (b) and (c) the crosses
         $(+)$ represent FC sites culled while the circles $(\circ)$ represent
         PC sites culled.  Each set of data points is taken from $1200$
         realizations of lattices of size $128 \times 128$.}
\label{fig:fcpc}
\vspace*{0.2in}
\end{figure}

The region near $\lambda = 0$ contributes to the behavior of the
random walker only for short time scales.  In this region the data for
IFC percolation peaks at $\lambda = 0$, similar to the way the solid curve 
for the uniform Euclidean case peaks in that region \cite{full}.  Also, 
as we iterate from ordinary to FC and then to IFC percolation, this 
closeness to the uniform Euclidean case improves.  Since the random walker 
perceives mostly the local structure of the cluster for short time scales, 
the local structure of the disordered cluster formed becomes increasingly
more Euclidean as we iterate from ordinary to FC and then to IFC 
percolation.  In particular, although critically disordered clusters 
formed in IFC percolation exhibit a fractal nature for long time scales 
they closely mimic the uniform Euclidean case for short time scales.

We further investigate the nature of the critically disordered 
clusters by culling either the fully-coordinated (FC) sites or the 
partially-coordinated (PC) sites.  FC sites are sites with four 
neighbors also occupied while PC sites have at least one neighbor 
that is empty.  Culling PC sites would produce clusters consisting 
only of the FC sites and thus leaving only the part of the cluster that 
is highly compact.

Shown in Fig.~\ref{fig:fcpc} are the eigenvalue spectrum when either FC or
PC sites are culled.  For ordinary percolation, when PC sites are culled
the eigenvalue spectrum is mostly zero.  However, when FC sites are culled
the eigenvalue spectrum still mimics the spectrum with no sites culled. 
This shows that PC sites are the dominant sites in clusters created from 
ordinary percolation in determining diffusion time scales, as they are
also dominant structurally \cite{cuansing01}.

For FC percolation, culling either FC or PC sites will produce eigenvalue 
spectra that behave in a similar fashion.  This also agrees with the 
previous result that neither FC nor PC sites dominate in the cluster's 
structure produced by FC percolation \cite{cuansing01}.

For IFC percolation, the eigenvalue spectrum when FC sites
are culled is mostly zero while the spectrum when PC sites are culled
produces most of the features for the original spectrum.  In particular,
consider the features near $\lambda = 0$ and near $\lambda = 1$.  The
spectrum when PC sites are culled behaves in a similar fashion to those
regions of the original spectrum where no culling is done (see Fig.
\ref{fig:spec}).  Thus, in IFC percolation FC sites are dominant not only
structurally but also in dictating the dynamics of the Brownian motion.

In summary, although FC and IFC percolation exhibit the same fractal
structure, short time diffusion characteristics become increasingly
Euclidean as we iterate the full coordination requirement.  Critically
disordered clusters produced from IFC percolation, in particular,
exhibit both a fractal structure for long time scales and a Euclidean
structure for short time scales.


\begin{thebibliography}{20}
\bibitem{full} In Euclidean (non-fractal) space, the density of states
         $\rho (E)$ for the corresponding classical vibration problem goes 
         as $\rho (E) \sim E^{d/2-1}$ in $d$ dimensions.  If we apply
         the correspondence $\lambda \sim e^{-E}$ (which is true
         at least for $\lambda$ close to $1$) as an approximation, then
         the density in $\lambda$ space should behave as
         $\rho (\lambda ) \sim \frac{1}{\lambda} |\ln \lambda |^{d/2-1}$.
         While this is not a rigorous argument, the $\frac{1}{\lambda}$
         dependence is consistent with the observed peak in $\rho (\lambda )$
         at $\lambda = 0$.
\end{thebibliography}
\end{document}